\documentclass[a4paper,11pt]{article}

\pdfoutput=1 
\usepackage{jcappub} 
\usepackage{eurosym}

\usepackage[T1]{fontenc} 

\usepackage{amssymb}
\usepackage{xcolor}
\usepackage[normalem]{ulem}
\usepackage{systeme}
\usepackage{amsmath}
\usepackage{bm}
\usepackage{verbatim}

\usepackage{hyperref}

\newcommand{\be}{\begin{equation}}
\newcommand{\ee}{\end{equation}}
\newcommand{\bea}{\begin{eqnarray}}
\newcommand{\eea}{\end{eqnarray}}

\def\d{{\rm d}}
\def\H{{\mathcal H}}
\def\n{\boldsymbol{n}}
\def\k{\boldsymbol{k}}
\def\v{\boldsymbol{v}}
\def\alm{a_{\ell m}}

\def\fsky{f_{\rm sky}}

\def\odm{{\Omega_{\rm cdm  0}}}
\def\ob{{\Omega_{\rm b 0}}}

\title{Constraining the growth rate by combining multiple future surveys}

\author[1]{Jan-Albert Viljoen,}
\author[1,2,3]{Jos\'e Fonseca,}
\author[1,4]{Roy Maartens}
\affiliation[1]{Department of Physics \& Astronomy, University of the Western Cape,\\Cape Town 7535, South Africa}
\affiliation[2]{INFN, Sezione di Padova, via Marzolo 8, I-35131, Padova, Italy}
\affiliation[3]{Dipartimento di Fisica e Astronomia ``G. Galilei'', Universit\`{a} degli Studi di Padova, Via Marzolo 8, 35131 Padova, Italy}
\affiliation[4]{Institute of Cosmology \& Gravitation, University of Portsmouth,  Portsmouth PO1 3FX, UK}

\emailAdd{javiljoen74@gmail.com}

\abstract{The growth rate of large-scale structure provides a powerful consistency test of the standard cosmological model and a probe of possible deviations from general relativity. We use a Fisher analysis to forecast constraints on the growth rate from a combination of next-generation spectroscopic surveys. In the overlap survey volumes, we use a multi-tracer analysis to significantly reduce the effect of cosmic variance. The non-overlap individual survey volumes are  included in the Fisher analysis in order to utilise the entire volume. We use the observed angular power spectrum, which naturally includes all wide-angle and lensing effects and circumvents the need for an Alcock-Paczynski correction. Cross correlations between redshift bins are included by using a novel technique to avoid computation of the sub-dominant contributions. Marginalising over the standard cosmological parameters, as well as the clustering bias in each redshift bin, we find that the precision on $\gamma$ improves  on the  best single-tracer precision by up to $\sim$50\%.}

\begin{document}
\maketitle
\flushbottom

\section{Introduction}

{General relativity and its classical modifications} (see e.g. the reviews \cite{Clifton:2011jh,Koyama:2015vza,Langlois:2018dxi,Frusciante:2019xia}) have distinctive effects on the clustering of galaxies and their peculiar velocities. Extracting the radial velocities of the matter distribution through redshift space distortions (RSD) of the 2-point galaxy correlations offers a powerful method of comparing different models of gravity and testing the consistency of the standard cosmological model. Such tests are mainly based on the growth rate $f$, or growth index $\gamma=\ln f/ \ln \Omega_{\rm m}$, and  require the redshift accuracy of spectroscopic galaxy surveys. {The state-of-the-art  measurement is from the extended Baryon Oscillation Spectroscopic Survey (eBOSS) Data Release 14 quasar (DR14Q) survey \cite{Zhao:2018jxv}, giving $\gamma=0.580\pm0.082$, which is consistent with the standard value $\gamma=0.545$.}

In the near future, various spectroscopic surveys will become operational, delivering an unprecedented view of the Universe, with exquisite precision. {Several papers have forecast how future surveys will constrain modifications of gravity via $f$ or $\gamma$ (see e.g. \cite{Bull:2015lja,Amendola:2016saw,Castorina:2019zho,Fonseca:2019qek}).} In addition to increasing the volume of observation, next-generation surveys will also use a range of wavelengths, creating complementary sets of dark matter tracers, whose cross-correlations can improve constraints and suppress some systematics \cite{Pourtsidou:2016dzn,Fonseca:2016xvi,Chen:2018qiu,Padmanabhan:2019xhc}.

{In fact, since the effect of RSD is similar to that of bias, one can use the multi-tracer technique \cite{Seljak:2008xr} to  minimise the effect of cosmic variance. Although the technique was initially proposed to measure local primordial non-Gaussianity without cosmic variance, its potential to better constrain the growth rate was shown by \cite{McDonald:2008sh}. Considerable effort has been applied to exploit information from multiple tracers with the goal of measuring RSD, including: improving power spectrum estimators \cite{Abramo:2015iga}; exploring the parameter space to better constrain the growth rate \cite{Abramo_2019,Boschetti:2020fxr}; and adding further velocity corrections \cite{Abramo:2017xnp}. An early  use of the multi-tracer technique in measuring the growth rate was in the GAMA survey \cite{Blake:2013nif}; for more recent applications to the 6dFGS survey, see \cite{Adams:2020dzw} and references therein.} 

Using the 3D power spectrum $P_g(\k,z)$ in Fourier space {has been the traditional approach to measure the growth rate} and allows one to cleanly separate the RSD effect via a Legendre multipole expansion, as  in the state-of-the-art eBOSS result \cite{Zhao:2018jxv}. Similarly, most forecasts for future surveys rely on the same analysis (e.g. \cite{Bull:2015lja,Amendola:2016saw,Pourtsidou:2016dzn,Castorina:2019zho}). 
The 2-point correlation function in redshift space, which is the Fourier partner of $P_g(\k,z)$, also allows for a simple multipole separation of RSD and has been applied to the eBOSS data in \cite{Tamone:2020qrl}.
Here we use the observed galaxy power spectrum in angular harmonic space, $C_{\ell}(z,z')$ \cite{Challinor:2011bk, Bonvin:2011bg,Bruni:2011ta, DiDio:2013sea, Tansella:2017rpi,Fonseca:2019qek}. This is the harmonic transform of the correlation function that is observed in redshift space; it not only enables the cross-correlation of redshift bins, but also naturally includes all wide-angle effects. 
In addition,  $C_{\ell}$ avoids the need for the Alcock-Paczynski correction, since the analysis of observations is performed directly in redshift space, without the need to assume a fiducial model in order to convert redshifts and angles to distances. It also naturally incorporates Doppler and lensing magnification effects on the correlations, which we include. The 2-point correlation function can also include all of these and other relativistic effects (see e.g. \cite{Tansella:2018sld}, which presents a fast code to compute the correlation function in the general case). Note that it requires an Alcock-Paczynski correction.

 The advantages of using $C_{\ell}$ do not come without drawbacks. In particular, we cannot cleanly separate out the RSD effect as in $P(\k,z)$ or the 2-point correlation function. Furthermore, we need to minimise the width of the window functions  in order to fully exploit the redshift accuracy of spectroscopic surveys,  leading to a high number of bins with rapidly growing computational cost from cross-bin correlations. Despite this, there have been many advances in using $C_{\ell}$ for analysis of survey data (see e.g. \cite{Asorey:2012rd,Campagne:2017xps,Assassi:2017lea,Gebhardt:2017chz,Camera:2018jys,Schoneberg:2018fis,Loureiro:2018qva, Alonso:2018jzx,Tanidis:2019teo}) and research in the field is ongoing. 
 
We use the Fisher forecast formalism for the angular power spectrum to estimate next-generation constraints on $\gamma$ using multiple tracers. 
For our forecasts, we  use the growth index $\gamma$, assumed to be constant.
Although this does not allow for scale-dependence of the growth rate (in common with most work on the growth rate),  it still delivers a consistency test of $\Lambda$CDM and standard dark energy models, and  a significant deviation of $\gamma$ from its 0.55 would indicate a breakdown of the standard model, due either to non-standard dark energy or modified gravity. 
Our parameter set includes the standard cosmological parameters as well as the clustering bias in each redshift window. We consider  upcoming spectroscopic surveys that use  galaxy counts (optical and near infra-red), and 21cm neutral hydrogen (HI) intensity mapping (IM). We use survey specifications that are similar to those planned for the DESI Bright Galaxy Sample \cite{Levi:2013gra}, for the Euclid H$\alpha$ survey \cite{Laureijs:2011gra}, and for the SKA HI IM surveys \cite{Bacon:2018dui}. This paper is based on our previous work \cite{Fonseca:2019qek}: here we expand the analysis to investigate the gain from combining these surveys, which were considered individually in  \cite{Fonseca:2019qek}. 

Using two distinctly different dark matter tracers that sample the same underlying density field enables us to significantly reduce the effect of cosmic variance. In addition, we include the  information from the remaining observed volume  by adding the Fisher matrices from non-overlap regions to Fisher information from the multi-tracer. In order to do this, we must assume that one can break the sky area into independent patches. The main implication of this is that one neglects modes above the size of the patches. Due to the large tomographic matrices we  break down the redshift range into independent subsurveys. All cross-bin correlations within each sub-survey are computed, but cross-correlations between subsurveys are neglected, as explained in more detail below.

We find that the errors on $\gamma$ (including Planck priors on standard cosmological parameters) from combining a high-$z$ SKA-like HI IM and a Euclid-like  H${\alpha}$ survey are $\sim$2.3\%. The combination of DESI-like BG and low-$z$ SKA HI IM surveys deliver $\sim$1.6\% precision. Combining all the information from high- and low-redshift surveys further improves the error on growth index to $\sim$1.3\%, which is an improvement of $\sim$55\% on constraints from the best independent survey.

{The paper is organised as follows. In \S\ref{sec:modgrav}, we briefly recap the growth index and the angular power spectrum. \S\ref{sec:comb_surv}  reviews the Fisher information matrix and how we include information from overlap and non-overlap volumes of the sky covered by the different surveys. We present our main results in \S\ref{sec:results} and conclude in \S\ref{sec:conclusion}.}

\section{Growth rate using the angular power spectrum} \label{sec:modgrav}

The irrotational peculiar velocity $v_i=\partial_i V$ is sourced by the comoving matter density contrast $\delta$ via the continuity equation:
\bea \label{eq:linvel}
\bm{\nabla}\cdot \v= \nabla^2 V=-f\,{\H}\, \delta\,, \quad f \equiv \frac{\d \ln D}{\d \ln a}\,,
\eea
where ${\cal H}=(\ln a)'$ is the conformal Hubble rate,  
and the growth factor $D$ is defined by $ \delta(a,\bm{k})=D(a)  \delta(1,\bm{k})$. We can parametrise $f$  in terms of {the matter density parameter} and growth index $\gamma$, as 
\be \label{eq:f_gamma}
f(a) = \Omega_{\rm m}(a)^\gamma \,,
\ee
where $\gamma=0.545$ gives a very good approximation to $f$ in $\Lambda$CDM and standard (non-clustering) dark energy models in general relativity.
{In order to increase the RSD signal in the angular power spectrum, we need to use finely-sliced tomographic information, leading to a very high number of redshift bins. In this case, it is better to constrain a single parameter $\gamma$ rather than $f$ in each redshift bin.}

{We use the angular power spectrum as an estimator of the matter fluctuations on the celestial sphere.} It is related to the two-point correlation function in redshift space by
\be
\big\langle \Delta(z_1,\n_1)\, \Delta(z_2,\n_2)\big\rangle = \sum_{\ell}\,{(2\ell+1) \over 4\pi}\, C_\ell(z_1,z_2)\,{\cal L}_\ell \big(\n_1\cdot\n_2 \big)\,,
\ee
where $\n$ is the unit direction of the source, $z$ is the observed redshift of the source, and ${\cal L}_\ell$ are Legendre polynomials. The observed redshift can be replaced by the background redshift  $z=a^{-1}-1$ 
at first order. The galaxy number density contrast or HI IM temperature contrast that is observed in redshift space is  \cite{Fonseca:2019qek}:
\bea
 \Delta &=&b\, \delta +~\,\mbox{RSD effect ~+ Doppler effect +  lensing effect} \notag\\
\label{obdel}
&=& b\, \delta-{1\over {\cal H}}\n \cdot \nabla \big(\v  \cdot \n \big)  +  A\big(\v  \cdot \n \big)+ (5s-2)\kappa\,,
\eea
where $b(z)$ is the Gaussian clustering bias,  $\kappa(z,\n)$ is the lensing magnification integrated along the line of sight, and the coefficient of the Doppler term is given by \cite{Fonseca:2019qek,Alonso:2015uua}
\bea
&& A~~ = b_e-\frac{\cal{ H}'}{{\cal H}^2} + \frac{5s-2}{\chi{\H}} -5s \,, \label{doppa}\\
&& s_{\rm HI}={2\over 5}\,,\qquad\qquad \qquad\qquad\quad~~ s_g ={2\over 5}\,{\Phi_{\rm cut}\over \bar{n}_g}\,, \label {magb}\\
&& b_e^{\rm HI}=- { \partial \ln \big[\big(1+z \big)^{-1}\H\, \bar{T}_{\rm HI}\big]\over \partial \ln (1+z)}  \,, ~~ b_e^g=- { \partial \ln \big[\big(1+z \big)^{-3} \bar{n}_g\big]\over \partial \ln (1+z)}\,. \label{evob}
\eea
Here $\chi(z)$ is the comoving line-of-sight distance, $s_g(z)$ is the galaxy magnification bias, and  $b_e(z)$ is known as the evolution bias. The background proper (i.e. physical) number density of galaxies above the flux cut is $\bar{n}_g$ (denoted $\bar{\cal N}_g$ in \cite{Fonseca:2019qek}) and the luminosity function  at the flux cut is $\Phi_{\rm cut}$. The background HI temperature is  $\bar{T}_{\rm HI}$.
We omit the Sachs-Wolfe, integrated Sachs-Wolfe and time-delay effects, which have a negligible impact on RSD measurements.

$\Delta$ can be expanded in spherical harmonics, with coefficients $a_{\ell m}$ that are assumed to be normally distributed.  Their covariance is 
\be
\big\langle a_{\ell m}(z_i)\, a^{*}_{\ell' m'}(z_j)\big\rangle = \delta_{\ell \ell'}\, \delta_{m m'}\, C_\ell(z_i,z_j)\,,
\ee
where $z_i$ are the redshift bin centres.
Following \cite{Challinor:2011bk}, we express the angular power spectrum in terms the primordial perturbations and the theoretical angular transfer functions $\Delta_\ell$ as 
\be \label{eq:clgeneral}
C_\ell ( z_i,z_j )=4\pi\!\!\int\!\!\d \ln k\, \Delta_\ell( z_i,k)\,\Delta_\ell (z_j,k)\, \mathcal P (k).
\ee
The power spectrum of primordial curvature perturbations is $\mathcal P (k)=A_s (k/k_s)^{n_s-1}$, where  $A_s$ is the amplitude (with pivot scale $k_s=0.05\,$Mpc$^{-1}$), and the spectral index is $n_s$. The  theoretical transfer function must be replaced by a windowed transfer function $\Delta_\ell^W(z_i,k)$, which is an integral over the $i$-bin weighted by a window function. Given the high-level redshift precision of the surveys that we consider, we use a top-hat window (smoothed at the edges to avoid numerical artifacts), as described in \cite{Fonseca:2019qek}.  
For more details, see \citep{Challinor:2011bk,Bonvin:2011bg,Bruni:2011ta,Jeong:2011as} for galaxy surveys and  \citep{Hall:2012wd,Fonseca:2018hsu} for maps of intensity.
\begin{figure}[!h]
\centering
\includegraphics[width=7.65cm]{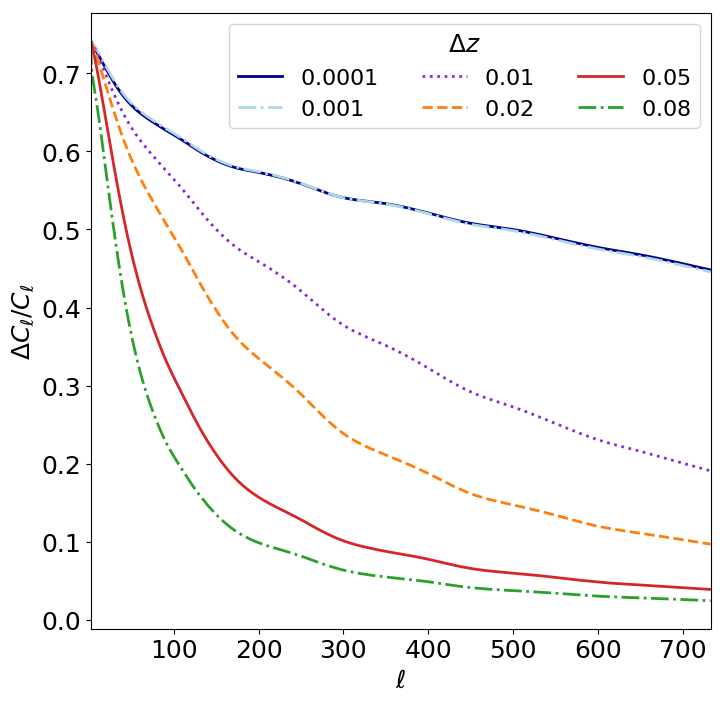}
\includegraphics[width=7.65cm]{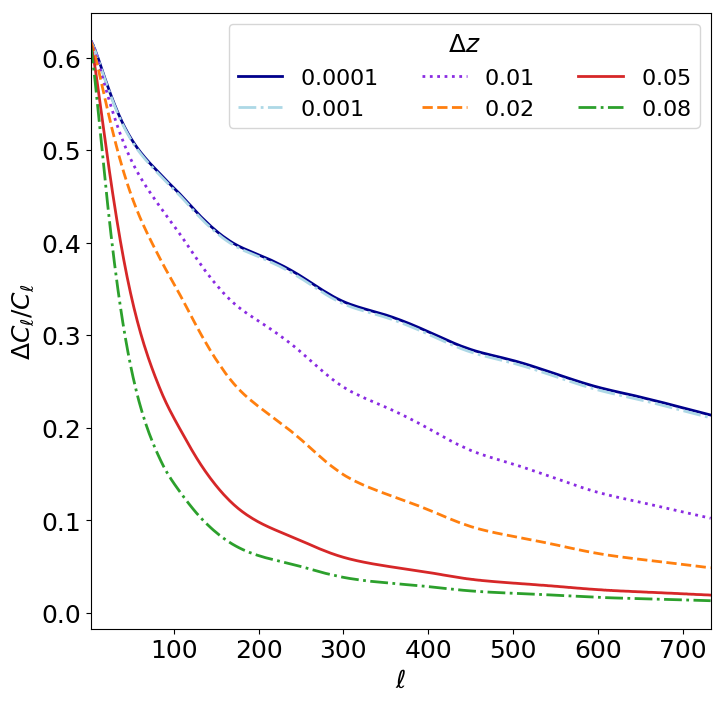}
\includegraphics[width=7.65cm]{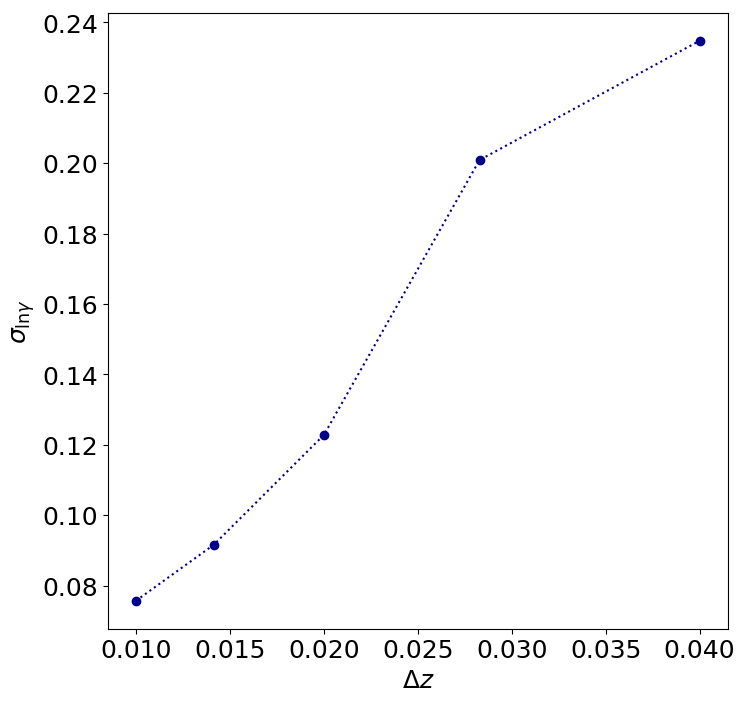}
\includegraphics[width=7.65cm]{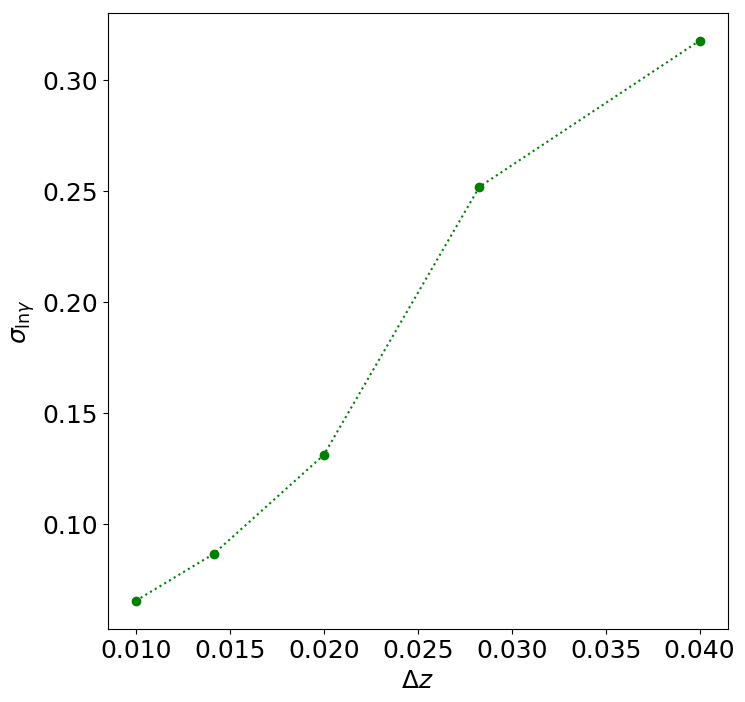}
\caption{{\em Top:} Fractional RSD contribution to $C_\ell$ as $\Delta z$ decreases, for HI IM ({\em left}) and H$\alpha$ galaxy ({\em right}) surveys at $z=1$. {\em Bottom:} For the same surveys in a fixed range $0.98<z<1.02$,
conditional fractional uncertainty on $\gamma$ as $\Delta z$ decreases. }
\label{fig:Delta_Cl}
\end{figure}

When measuring RSD in the angular power spectrum, the RSD signal to noise increases as the redshift bin width $\Delta z$ is decreased \cite{Fonseca:2019qek}.  As shown in Figure~\ref{fig:Delta_Cl} (top),
the fractional contribution of RSD\footnote{$C_{\ell}\,{\rm (total)}$ includes the Doppler and lensing contributions, but they are dwarfed by RSD.}, $[C_{\ell}\,{\rm (total)}-C_{\ell}\,\mbox{(density only)}]/C_{\ell}\,{\rm (total)}$, 
grows as $\Delta z$ shrinks. Although the noise increases at the same time, the number of auto- and cross-bin correlations also increases to compensate the effect of the noise. The result is a growth in signal-to-noise with decreasing $\Delta z$, as illustrated in the bottom panels. This means that the best constraints from a survey will come from using $\Delta z$ equal to the limiting redshift resolution of the telescope. For next-generation surveys, this is typically $\sim$$10^{-4}$, which creates a major computational obstacle. A compromise is to use $\Delta z=0.01$.

\section{Combining multiple surveys} \label{sec:comb_surv}

In our previous work \cite{Fonseca:2019qek} we forecasted precision on $\gamma$ from individual  next-generation spectroscopic surveys. A natural extension is to combine different surveys and take advantage of the cross-correlations between tracers to improve the statistical power. Each survey scans a particular sky area and redshift range, which does not necessarily overlap with another survey. A first approach is to combine surveys via a joint analysis. This is a good approach when we consider different cosmological probes, for example, joining the information from SNIa supernovae and from the CMB. When we consider different galaxy surveys, we can no longer do joint analysis at the posterior level. 

A possible approach  is the multi-tracer technique \cite{Seljak:2008xr}, which requires a perfectly overlapping volume, i.e., the same redshift range and sky area. This was applied in \cite{Alonso:2015sfa,Fonseca:2015laa}, using the angular power spectra, to constrain local-type primordial non-Gaussianity, leading to significant improvements over single-tracer constraints as a result of the suppression of cosmic variance. Primordial non-Gaussianity in the power spectrum is an ultra-large scale effect and is therefore heavily impacted by cosmic variance. As a consequence, a smaller overlap volume still produces better results than a simple combination (neglecting the cross-tracer correlations) of the full larger volume of each individual tracer. Since RSD measurements do not rely on ultra-large scales, the gain from the multi-tracer is lower
and we need to combine information from non-overlap volumes with the multi-tracer information.

Using the  $\alm$  as the observable, the  Fisher matrix is given by \citep{Tegmark:1996bz}
\bea \label{eq:fishercl}
{ F}_{\alpha \beta}=\sum_{\ell_{\rm min}}^{\ell_{\rm max}} \frac{(2\ell+1)}{2 }f_{\rm sky}\,{\rm Tr}\Big[ \big(\partial_{\alpha} \bm{C}_{\ell}\big)\, \bm{\Gamma}_\ell^{-1} \big(\partial_{\beta}\bm{C}_{\ell}\big)\,\bm{\Gamma}_\ell^{-1}\Big]\,,
\eea
where the trace over the matrix product is effectively a sum over all auto- and cross-bin correlations in the redshift range of the survey.
Here $\partial_\alpha=\partial/\partial \vartheta_\alpha$ where $\vartheta_\alpha$ are the parameters, and the matrices are
\be
\bm{C}_{\ell}=\big[ C_\ell(z_i,z_j)\big],\quad  \bm{\Gamma}_\ell =\big[ \Gamma_\ell(z_i,z_j)\big]=\bm{C}_\ell+\bm{\mathcal N}_\ell,
\ee
where $\bm{\mathcal N}_\ell$ is the noise. The maximum angular scale included is given by $\ell_{\rm min}$, which is determined by the sky fraction: $\ell_{\rm min}={\rm int}(\pi/\sqrt{\Omega})+1$ where $\fsky=\Omega/4\pi$. There is loss of ultra-large scale modes due to systematics (e.g. from foreground cleaning of IM or dust extinction from the galaxy modulating the threshold flux limit), and therefore $\ell_{\rm min}$ may need to be increased above this value. In order to take account of this, we impose $\ell_{\rm min}\geq 5$.
 The minimum angular scale is given by $\ell_{\rm max}$. We  impose the cut  proposed in   \cite{Fonseca:2019qek} {to include only linear scales}. 

Following  \cite{Fonseca:2019qek}, we apply two techniques in order to achieve manageable numerical computations when redshift bin widths are $\sim$0.01, giving $O(100)$ bins and the same number of bias nuisance parameters: 
\begin{itemize}
\item
We divide the redshift range into subsurveys and in each subsurvey, all auto- and cross-correlations are computed. Cross-correlations between subsurveys are omitted. Provided that the subsurveys are of sufficient width (typically $\gtrsim 0.1$), this technique
has been shown in  \cite{Camera:2018jys} to include the dominant cross-bin correlations with little loss of information and negligible bias on cosmological parameter measurements. 
\item
We reduce the number of parameters by marginalising out the $O(100)$ bias nuisance parameters $b(z_i)$, leaving $\gamma$ and 6 standard cosmological parameters (for which we use the fiducial values and Gaussian priors from Planck 2018 \cite{Aghanim:2018eyx}):
\be\label{para}
 \vartheta_\alpha = \big( \ln \gamma, \ln A_s, \ln n_s,\ln\odm,\ln \ob, w, \ln H_0  \big) \,.
\ee  
Only the 7 cosmological parameters \eqref{para} are present in all the Fisher sub-matrices.
The constraints from  subsurvey $s_I$ are computed using \eqref{eq:fishercl}, and then we add the Fisher information matrices from all subsurveys:
\be
{ F}_{\alpha \beta}=\sum_I { F}_{\alpha \beta} (s_I)\,.\label{eq:Fsub}
\ee
\end{itemize}

We can generalise  to the multi-tracer combination of two\footnote{
Note that one can generalise this to more than 2 tracers, e.g. to find the internal covariance of experiments with multiple probes \cite{Krause_2017, Ballardini:2019wxj}.} 
dark matter tracers $A$ and $B$, with the same sky area $\Omega_{AB}$,  the same redshift range and the same redshift binning, by using the combined matrix  \cite{Alonso:2015sfa,Fonseca:2015laa}:
\bea
\bm{ C}_\ell
=
\begin{bmatrix}
  { C}^{AA}_\ell ( z_i,  z_j) &  ~~~{C}^{AB}_\ell ( z_i,  z_j) \\ &&\\
  {C}^{BA}_\ell ( z_i,  z_j) &  ~~~{C}^{BB}_\ell ( z_i,  z_j)
\end{bmatrix}\,, \label{eq:cov_overlap}
\eea
in \eqref{eq:fishercl}. Similarly to the single-tracer case,
we apply the subsurvey division of the common redshift range, marginalise out the bias parameters $b^A(z_i)$ and $b^B(z_i)$, and add the subsurvey matrices
  to produce a multi-tracer Fisher matrix ${F}^{AB}_{\alpha\beta}(\mbox{overlap})$ on the overlap volume, computed using \eqref{eq:fishercl}  and \eqref{eq:cov_overlap}.

In general, surveys $A$ and $B$ will not have the same sky area and the same redshift ranges. In this case, there is additional information in the non-overlap volumes of the two surveys $A$ and $B$. These non-overlap volumes include in general two contributions:
\begin{itemize}
\item 
the non-overlap parts of each sky area, $\Omega_A-\Omega_{AB}$ and 
$\Omega_B-\Omega_{AB}$, across the full redshift range for each survey, $z_{\rm max}^A -z_{\rm min}^A$ and $z_{\rm max}^B -z_{\rm min}^B$; 
\item
the overlap sky area $\Omega_{AB}$, across the non-overlap parts of the redshift ranges.
\end{itemize}
The non-overlap volumes are processed in the same way as above -- divide into subsurveys, marginalise out the bias parameters, and add the subsurvey Fisher matrices. This produces two non-overlap Fisher matrices, $F^A_{\alpha\beta}(\mbox{non-overlap})$ and $F^B_{\alpha\beta}(\mbox{non-overlap})$, which are then added to the overlap multi-tracer Fisher matrix to produce the total Fisher matrix: 
\be \label{total}
F_{\alpha\beta}(\mbox{total}) = F^{AB}_{\alpha\beta}(\mbox{overlap}) + F^A_{\alpha\beta}(\mbox{non-overlap}) + F^B_{\alpha\beta}(\mbox{non-overlap})\,.
\ee

The noise matrix in \eqref{eq:fishercl} is given by shot noise for galaxy counts and instrumental noise for IM: 
\bea \label{slg}
{\cal N}^{g}_\ell(z_i,z_j) & =&\frac1{ {N_\Omega} (z_i)}\, \delta_{ij}\,,\\
 \label{eq:inst_noise_hiim}
{\cal N}^{\rm HI}_\ell(z_i,z_j)  &=&\frac{4\pi\,f_{\rm sky}}{2\, N_{\rm d}\, \Delta \nu(z_i)\, t_{\rm tot}} \,T_{\rm sys}(z_i)^2 \,
\, \delta_{ij}\,.
\eea 
For galaxy surveys, ${N_\Omega} (z_i)$ is  the average angular density of sources in the bin.
For IM, in single-dish mode, 
  $T_{\rm sys}$ is a sum of the  temperatures of dish receivers and the sky, $N_{\rm d}$ is the number of dishes, $\Delta \nu$ is the {bin size in frequency}  and $t_{\rm tot}$ is the total integration time. 
The shot noise for IM is much smaller than the instrumental noise  \eqref{eq:inst_noise_hiim} on the linear scales considered here: see e.g.
 \cite{Chang:2007xk,Gong:2011qf,Castorina:2016bfm,Villaescusa-Navarro:2018vsg} and Appendix \ref{apx:Xshot}.
We also neglect cross-shot noise contributions from the correlations between galaxy surveys and HI IM, since they are small \cite{Fonseca:2015laa}. In Appendix \ref{apx:Xshot}, we estimate this cross-shot noise and argue why it should be negligible.

\section{Prospects from next-generation spectroscopic galaxy surveys}\label{sec:results}

We consider near future surveys, such as SKA Phase 1 IM and  galaxy surveys with specifications similar to those in the literature for DESI and Euclid. The various survey specifications and noise assumptions we use in this paper are given in more detail in our previous work \cite{Fonseca:2019qek}. We summarise the basic experimental details in Table~\ref{tab:survdetails} and below we give the main assumptions for the astrophysical details of each survey. 

For a survey similar to the one planned for the Bright Galaxy Sample (BGS) of DESI, we use the fits to simulations from \cite{Aghamousa:2016zmz}:
\bea
N^{\rm BGS}_g &=& 6.0\times 10^3 \Big( \frac z{0.28}\Big)^{0.91} \exp\Big[-\Big({z\over 0.28}\Big)^{2.56}\Big] ~~ {\rm gal}/\deg^{2}\,, \label{eq:nbgs} \\
b^{\rm BGS}_g&=&0.99+0.73 \,z-1.29\, z^2+10.21\, z^3 \label{eq:bbgs}\,.
\eea
Here $N_g$ is related to $\bar{n}_g$ as $N_g=(1+z)^{-4}c\H^{-1}\chi^2\,\bar{n}_g$. Then we can use \eqref{evob}  to compute $b_e^{\rm BGS}$. Since the BGS  is a low-redshift sample, the lensing contribution  is very small and we can safely neglect $\kappa$ in \eqref{obdel}. In the Doppler term \eqref{doppa} for the BGS, we set $s=0$ since the magnification effect is negligible.

For an  H$\alpha$ spectroscopic survey similar to that planned for Euclid, we update our previous specifications in \cite{Fonseca:2019qek}, in light of \cite{Blanchard:2019oqi}.  We use the  Model 3 luminosity function of \cite{Pozzetti:2016cch} to compute the number density, and then compute the magnification bias \eqref{magb} and evolution bias \eqref{evob}. The clustering bias model for the H$\alpha$ sample is based on recent estimates by \cite{Merson:2019vfr}. Then the astrophysical details for the H$\alpha$ survey are
\bea
N^{{\rm H} \alpha}_g &=& z^{0.83} \exp\left(9.34+0.19 z-1.92 z^2+0.63 z^3-0.07z^4\right) ~~ {\rm gal}/\deg^2 \,, \label{eq:nHag}\\
b^{\rm H\alpha}_{g} &=& 0.7 (1+z) \,,\\
s^{\rm H\alpha}_g&=& 0.27+0.62\, z- 0.03 \,z^2 - 0.07\, z^3-0.02\, z^4 \label{eq:bHag}\,.
\eea

In the case of HI IM (Band 1 and Band 2), for the background HI temperature and the HI bias  we use \cite{Fonseca:2019qek}:
\bea
\bar{T}_{\rm HI}(z)&=&0.056 + 0.232 \,z - 0.024\, z^2\,, \\
b_{\rm HI}(z)&=&0.667 + 0.178\, z + 0.050\, z^2\,.
\eea
 HI IM does not have a lensing correction to the angular power spectrum at linear order, as reflected in \eqref{magb}. The evolution bias is given by \eqref{evob}. We include a Gaussian beam in $C^{\rm HI}_\ell$, to account for the optics of the dishes (see \cite{Fonseca:2019qek}).
 
 Note that in all surveys, the clustering bias model simply provides a fiducial value $b^A(z_i)$ in each bin, and we marginalise over the uncertainty in the bias. 

\begin{table}[!h]
\caption{\label{tab:survdetails} Volumes of next-generation spectroscopic surveys.}
\centering
\begin{tabular}{llcl}
\\ \hline\hline
Experiment & Tracer &$\Omega_{\rm sky}$& Redshift \\
 & &$[10^3\deg^2]$ & range \\
\hline
{SKA1 IM2} & {HI IM} & 20 & 0.1--0.58 \\
{SKA1 IM1} & {HI IM} & 20 & 0.35--3.06 \\
Euclid-like & H$\alpha$ galaxies & 15 & 0.9--1.8 \\
DESI-like & Bright galaxies & 15 & 0.1--0.6 \\
\hline
\hline
\end{tabular}
\end{table}
\begin{table}[h]
\caption{\label{tab:survs_comb} Overlap and non-overlap sky areas for the low- and high-$z$ combinations.}
\centering
\begin{tabular}{lccccc}
\\ \hline\hline
 & Tracer A & Tracer B & $\Omega_{A}-\Omega_{AB}$ & $\Omega_{B}-\Omega_{AB}$ & $\Omega_{AB}$ \\
 &        &        & $[10^3\deg^2]$ & $[10^3\deg^2]$ & $[10^3\deg^2]$\\
\hline
Low-$z$ &SKA1 IM2 & BGS & 10 & 5 & 10 \\
High-$z$ &SKA1 IM1 & H$\alpha$ & 10 & 5 & 10 \\
\hline
\hline
\end{tabular}
\end{table}

Our goal is to combine these surveys to find the prospects of testing gravity  in the near future, using linear scales. Table~\ref{tab:survdetails} shows that there are significant overlaps in the low- and high-redshift ranges, suggesting a multi-tracer combination of IM2 with BGS and another of IM1 with H$\alpha$. 
 In Table \ref{tab:survs_comb} we summarise what is assumed for the overlap area, $\Omega_{AB}$, which then gives the non-overlap areas, $\Omega_A-\Omega_{AB}$ and 
$\Omega_B-\Omega_{AB}$. We fix the HI IM instrumental noise  \eqref{eq:inst_noise_hiim} by fixing the scanning ratio, i.e, the sky area over time. This implies that the observational time $t_{\rm tot}$, has to be adjusted proportionally to the reduction in sky area. Further details on subtleties in the SKA1 noise are given in \cite{Fonseca:2019qek}.

For the low redshift combination there is a good overlap in the redshift range, but for the high redshift case IM1 extends well beyond the H$\alpha$ range. In practice above $z=1.8$ and below $z=0.9$ we only obtain constraints from HI IM, although we still add this information to the overall constraints as in \eqref{total}.  

We fix the subsurveys to have 20 redshift bins of width 0.01. Note that subsurveys at the edges of the redshift range may  have less than 20 redshift bins.

 \subsection*{Results}

\begin{table}[!h]
\caption{Normalised errors on $\gamma$.}
\label{tab:marginal_gamma}
\centering
\begin{tabular}{llcc}
\\ \hline\hline
 &Survey &  {\centering $\sigma_{\ln \gamma}\;(\%)$}\\
\hline
Low redshift &BGS & 4.7 \\
&SKA1 IM2 & 2.9\\
&Combined total: IM2+BGS & 1.6 \\
\hline
High redshift &H$\alpha$ survey & 4.0 \\ 
&SKA1 IM1 & 3.8\\
&Combined total: IM1+H$\alpha$ & 2.3 \\
\hline
&&\\
Low + High redshift & Combined total: IM2+BGS+IM1+H$\alpha$  & 1.3 \\
\hline\hline
\end{tabular}
\end{table}

\noindent Table \ref{tab:marginal_gamma} summarises our results. The single-tracer errors compare well with our previous results \cite{Fonseca:2019qek}, except for the H$\alpha$ sample, whose specifications we have updated. There is a very small difference for the other surveys since we use narrower subsurveys than   \cite{Fonseca:2019qek}. 

The new results are for the combined totals of surveys, i.e., using multi-tracer in the overlap volume and adding single-tracer in the non-overlap volumes. As expected, the combination at low redshift has more constraining power than the one at high redshift, given that $f$ and $\Omega_{\rm m}$ are tending to 1 at higher redshifts. 
 
\begin{figure}[!h]
\centering
\includegraphics[width=7.65cm]{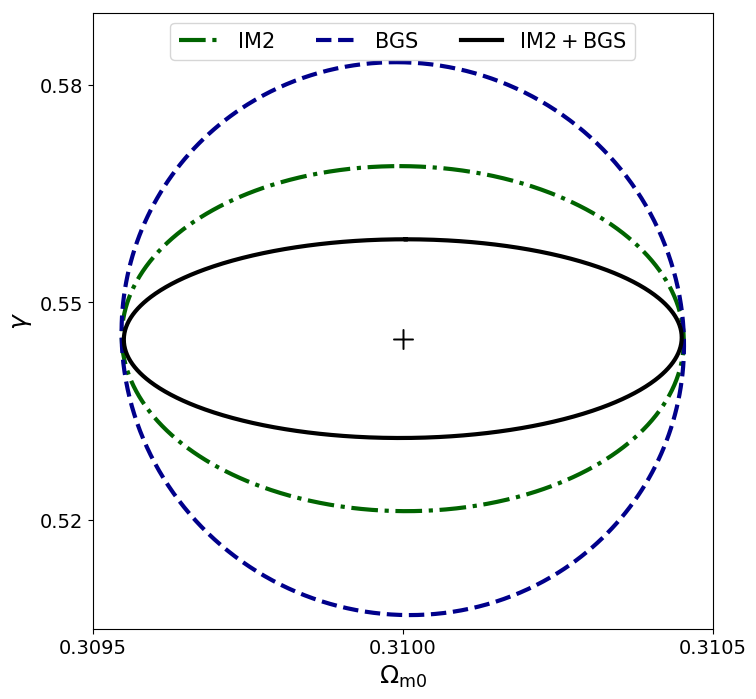}
\includegraphics[width=7.65cm]{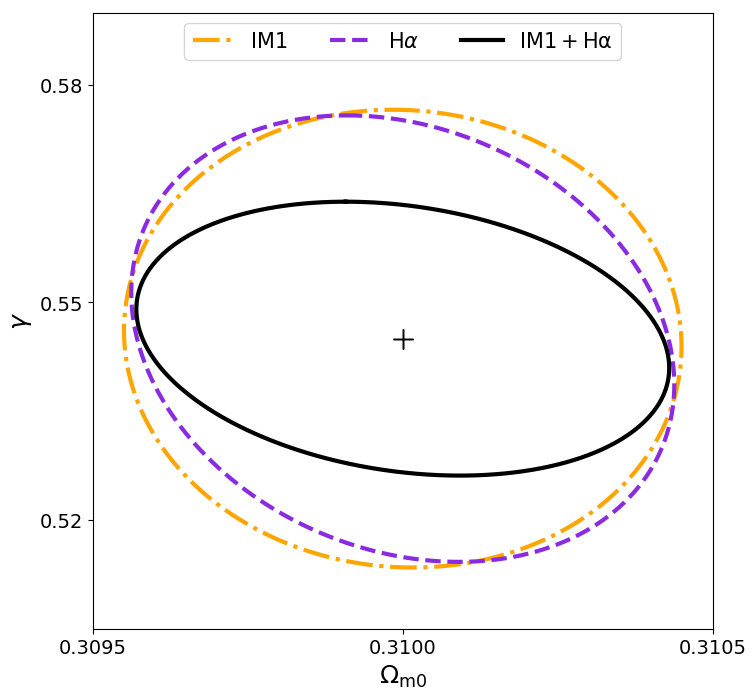}
\caption{Marginal 1$\sigma$ contours for matter density and growth index: low-$z$ surveys ({\em left}); high-$z$ surveys ({\em right}). Solid black contours denote the combined total, as in \eqref{total}, and the + indicates fiducial values.}
\label{fig:contour_errorsA}
\end{figure}

The constraints on $\gamma$ are degenerate with $\Omega_{\rm m0}$ by \eqref{eq:f_gamma}. In Figure \ref{fig:contour_errorsA} we plot the 1$\sigma$ contours for the low-$z$ (left) and high-$z$ (right) surveys as well as their combinations. 
Note that since $\Omega_{\rm m0}=\Omega_{\rm cdm0}+\Omega_{\rm b0}$, we had to transform our Fisher matrix adding the constrains from both parameters (see \cite{2009arXiv0906.4123C}).  

At higher redshift we find that there is still substantial  information outside the overlap volume of IM1 and H$\alpha$ surveys. In Appendix \ref{apx:comp_MT_allarea} we repeat our forecasts in the traditional multi-tracer analysis where one only considers the overlap volume. For the low-$z$ combination the degradation in precision is not strong, but for the high-$z$ case the overlap volume alone is not even competitive with the  single tracer constraints.

\begin{figure}
\centering
\includegraphics[scale=0.41]{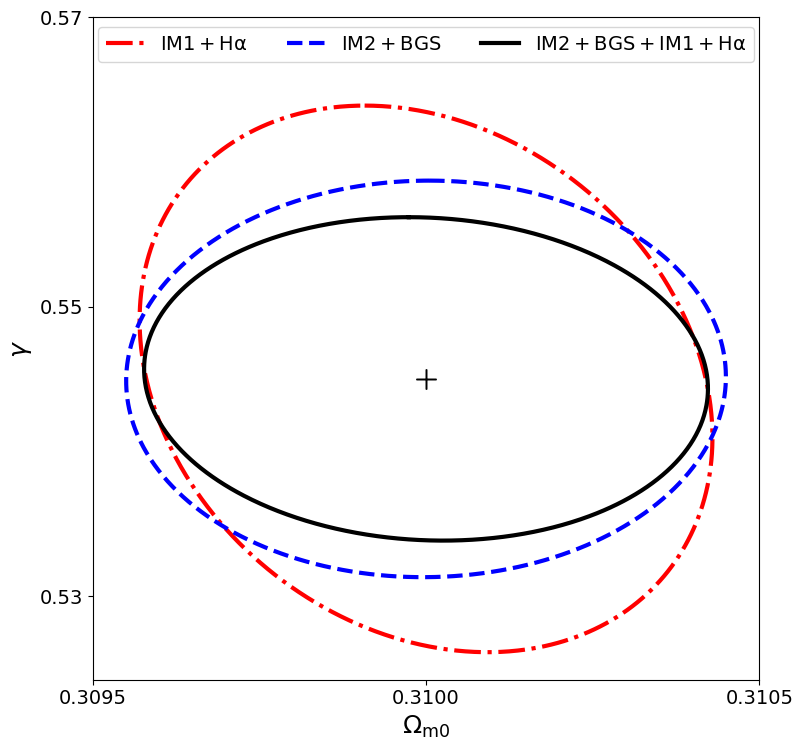}
\caption{\label{fig:contour_total} Marginal {1$\sigma$} contour plots for the low- and high-$z$ combined totals and for the overall combined total (solid black).}
\label{fig:contour_errors_tot}
\end{figure}

By combining surveys and utilising the full observed volume, we  find better results than the best single-tracer survey result. It is therefore natural to extend the combination by a further step --  adding the combined totals  from low and high redshift, assuming that they too are independent. There is a caveat: in order to avoid double-counting of the IM signal, we remove from the high-$z$ combination the contribution with $z<0.6$. 
The result is given in the last row of Table~\ref{tab:marginal_gamma} ($\sigma_\gamma/\gamma = 1.3\%$) and Figure~\ref{fig:contour_errors_tot} displays the 1$\sigma$ contours\footnote{Note that solid black contour is not exactly the combination of the red dot-dashed  and blue dashed contours, since we removed some IM1 bins to avoid double-counting.}.

\section{Conclusions} \label{sec:conclusion}

{We investigated the constraints on the growth rate parameter $\gamma$ that can be expected from roughly contemporaneous next-generation spectroscopic surveys, using only linear scales. Our goal was not to forecast individual survey constraints, which was done in our previous work \cite{Fonseca:2019qek}. Here we wanted to include all possible information from these surveys, using multi-tracer cross-correlations on overlap volumes and single-tracer correlations on non-overlap volumes. To do this we assumed that different patches of the sky are independent and we only included modes that are contained within each patch.}

We used the growth index $\gamma$ rather than the growth rate $f$, since it is redshift-independent and therefore better suited to surveys with very high numbers of redshift bins. Although the $\gamma$ parametrisation is not valid for scale-dependent modifications of gravity, it still provides a test of the standard cosmological model and a probe of possible deviations from general relativity.

We used the observed angular power spectrum $C_\ell$, whose key advantages include: it incorporates the redshift evolution of all cosmological, astrophysical, and noise variables; it does not impose a flat-sky approximation but naturally incorporates all wide-angle effects; Doppler and lensing magnification corrections to the 2-point correlations are also naturally included. Furthermore, since it is directly observable, the angular power spectrum of the data requires no fiducial model and therefore no Alcock-Paczynski correction is needed. These advantages over the Fourier power spectrum $P_g$ (which is not a direct observable) come with a price.  $P_g$ delivers a clean separation of the RSD effect, unlike $C_\ell$. In addition, there are computational challenges in extracting maximal information from $C_\ell$. In particular, performing all cross-bin correlations becomes increasingly difficult for the very thin bins that are needed to enhance the RSD signal. These computational challenges can be mitigated by a `hybrid' method which divides the full redshift range into independent subsurveys. The full range of  auto- and cross-bin correlations are computed only within each subsurvey.

We marginalised over the standard cosmological parameters, as well as  the clustering bias in each redshift bin. We used information only from linear scales. Our main results are shown in Table~\ref{tab:marginal_gamma} and in the contour plots of Figures~\ref{fig:contour_errorsA} and \ref{fig:contour_total}. The best marginal constraints on $\gamma$ are $\sim$1.6 and $\sim$2.3\% for combinations of low- and high-$z$ surveys respectively.  These are $\sim$45\% tighter than the best independent survey. If we take the further step of combining the low- and high-$z$ combinations,  we find a precision of $1.3\%$, which is $\sim$55\% better  than the best single-tracer. 

In summary, combining the information from appropriate near-future spectroscopic surveys -- via the multi-tracer technique in the overlap volumes and the single-tracer in non-overlap volumes -- will significantly improve constraints of the growth rate of large-scale structure, without using more observational resources. 

\acknowledgments
We thank Mario Ballardini for useful discussion and comments on this work, part of which occured at the University of Bologna, supported by the Italian South-African Research Programme (ISARP). We also thank Dionysis Karagiannis for helpful comments.
JF is supported by the University of Padova under the STARS Grants programme {\em CoGITO: Cosmology beyond Gaussianity, Inference, Theory and Observations.} {JF also thanks the University of the Western Cape for supporting a visit during which parts of this work were developed.}
JV and RM are supported by the South African Radio Astronomy Observatory and the National Research Foundation (Grant No. 75415). RM is also supported by the UK Science \& Technology Facilities Council (Grant ST/N000668/1).
This work  made use of the South African Centre for High Performance Computing, under the project {\em Cosmology with Radio Telescopes,} ASTRO-0945
 
\newpage

\appendix

\section{Shot noise and cross-shot noise}\label{apx:Xshot}

For correlations between tracers one expects an overlap in the dark matter halos seen by both. An exception is for example the red and blue galaxies in photometric galaxy surveys, which by selection are disjoint tracers of the dark matter. When we consider HI IM, all halos in a {voxel} that contain HI will contribute to the integrated temperature observed in the {voxel}. Some of these halos, especially the most massive ones, will host emission line galaxies which appear in spectroscopic galaxy surveys such as the DESI-like BGS survey. These overlap halos will induce a shot-noise term contribution in the cross-correlation. 

The comoving HI density is given by 
\be
\rho_{\rm HI} =\int^{M_{\rm HI}^{\rm max}}_{M_{\rm HI}^{\rm min}} {\rm d}M\, {n_h(M)}\, M_{\rm HI}(M)\,,
\ee
where $n_h$ is the halo mass function, $M_{\rm HI}$ is the mass of HI in a halo of mass $M$, and we take $M_{\rm HI}(M)\propto M^{0.6}$ \cite{Santos:2015gra}. 
The HI shot noise power spectrum is   \cite{Castorina:2016bfm,Villaescusa-Navarro:2018vsg}
\be
{{P}^{\rm HI}_{\rm sn}= {\bar{T}_{\rm HI}^2 \over \rho_{\rm HI}^2} \int^{M_{\rm HI}^{\rm max}}_{M_{\rm HI}^{\rm min}}  {\rm d}M\, n_h(M)\, M_{\rm HI}(M)^2\,,}
\ee
and the galaxy shot noise power spectrum is $P^g_{\rm sn}=(a^3\bar{n}_{g})^{-1}$.
The cross-shot noise power spectrum can be estimated as \cite{Fonseca:2015laa} 
\be \label{eq:Xshot}
{{P}^{\times}_{\rm sn}={1\over a^3\bar{n}_{g}}\, \frac{\bar{T}_{\rm HI}}{\rho_{\rm HI} } \int^{M^{\rm max}}_{M^{\rm min}}  {\rm d}M\, n_h(M)\, M_{\rm HI}(M)\,\Theta(M)\,,}
\ee
where $\Theta$ is a weighting to account for the fraction of halos that are present in both samples.
For simplicity, we dropped the redshift dependence, and in what follows we neglect the width of the bins and just take their central values. In the absence of an exact model, one can approximate  that all halos within a  mass range {$M^{\rm min}\leq M \leq M^{\rm max}$} overlap and set $\Theta(M)=1$ in this range and zero elsewhere. This implicitly assumes that all halos within this mass range {have HI and host a bright galaxy,} which is incorrect. However, assuming $\Theta(M)=1$ leads to an over-estimation of the noise contribution within the mass range. 
\begin{figure}
\centering
\includegraphics[width=7.6cm]{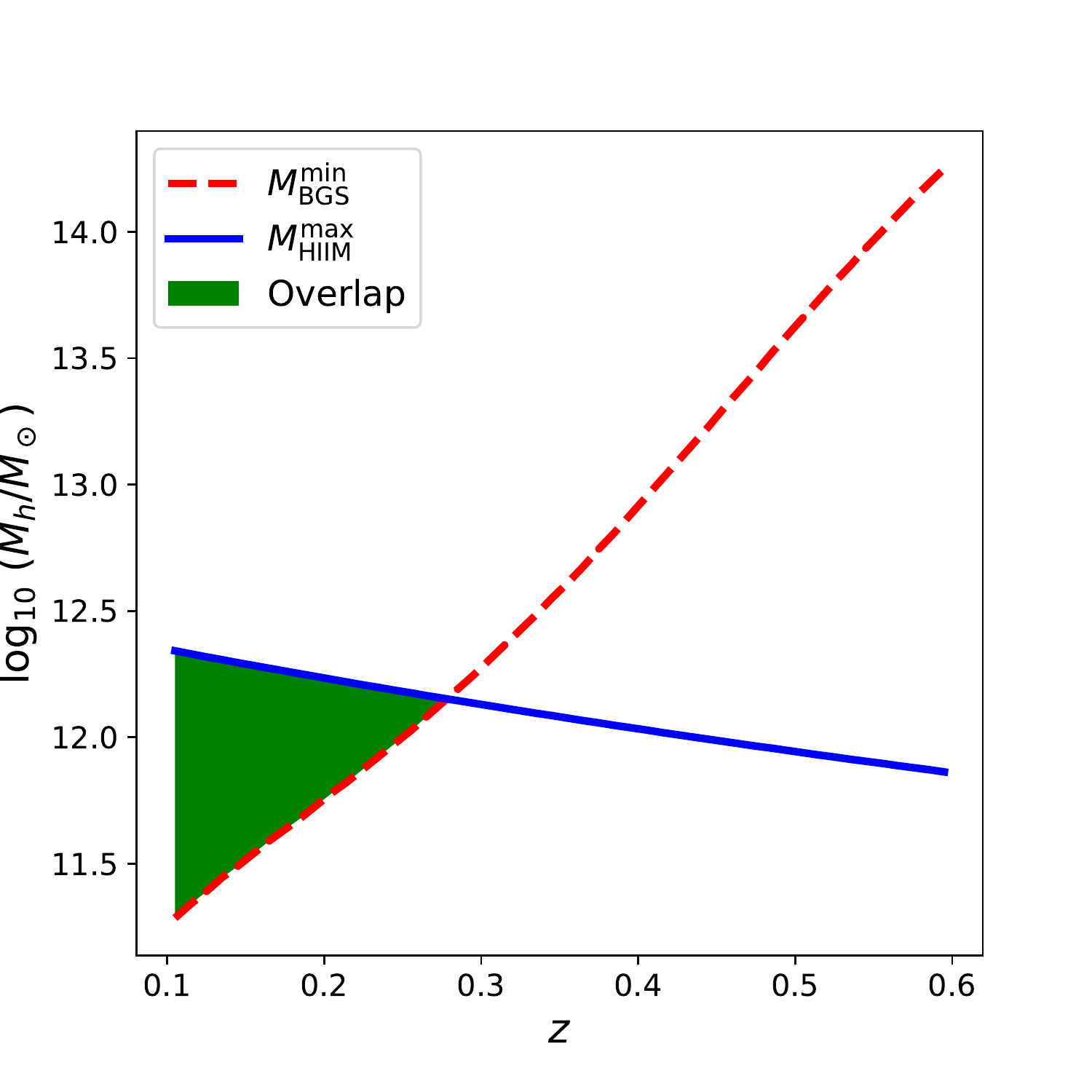}
\includegraphics[width=7.6cm]{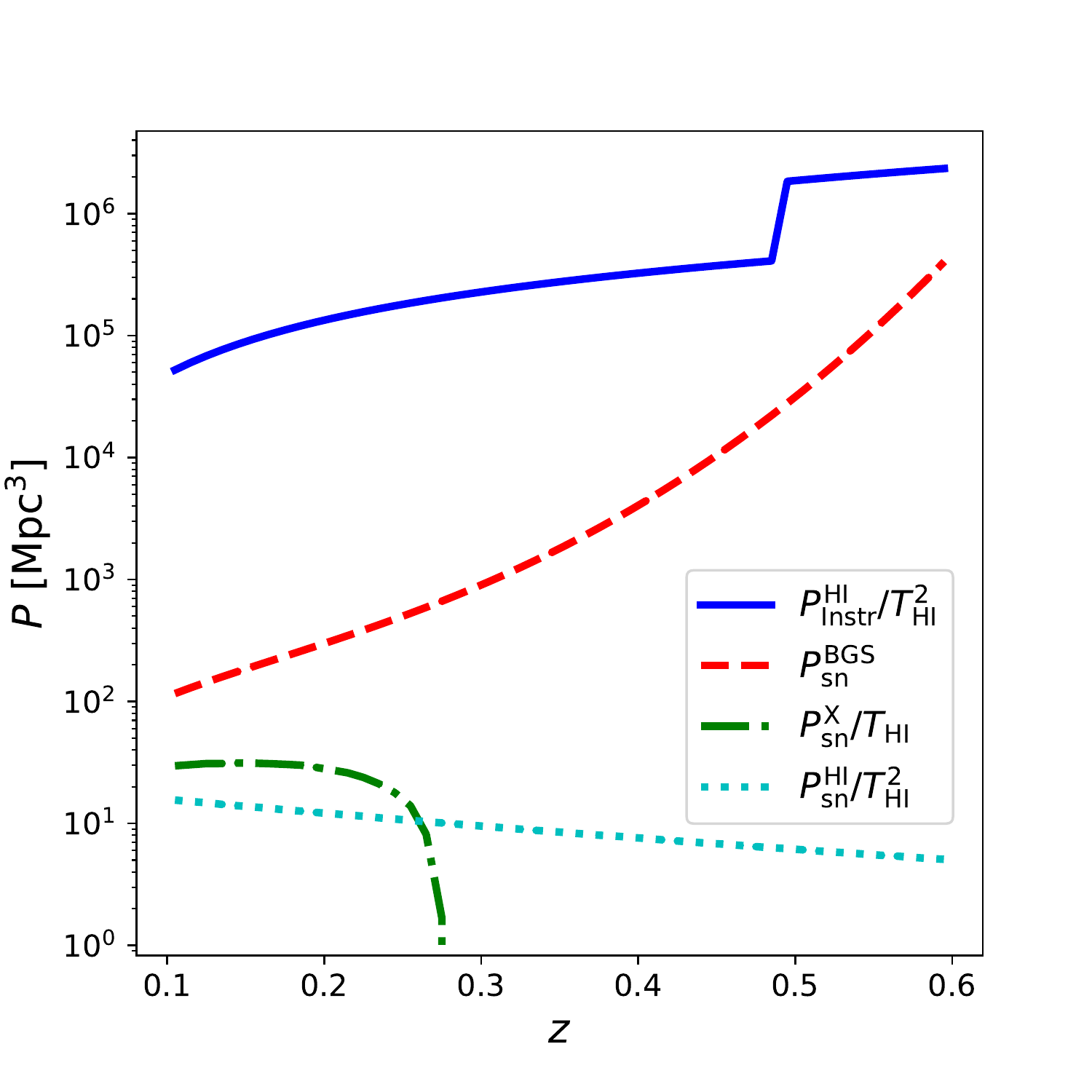}
\caption{{\em Left:} Mass overlap between HI {IM} and BGS {halos}. 
{\em Right:} Shot-noise power spectra for HI IM and BGS, together with the over-estimate of their cross-shot noise. The instrumental-noise power spectrum for HI IM is also shown. }
\label{fig:Xnoise_mass_range}
\end{figure}

In order to find  $M^{\rm min}$ and $M^{\rm max}$ {in \eqref{eq:Xshot}}, we need to estimate the halo mass ranges of each survey. The BGS survey will cover a range of higher mass halos, while the HI IM surveys cover lower mass halos which are small enough to contain neutral hydrogen. For the BGS survey, we use abundance matching between the number of galaxies in a bin and the expected number from the halo mass function, i.e.,
\be
{a^3\bar{n}_{g}}= (1+z) \frac \H{c\chi^2} \,N_{g}=\int^{\infty}_{M_{g}^{\rm min}} {\rm d}M \,n_h(M)\,. 
\ee
In this approximation we assume that all massive halos will host a bright galaxy, which is not necessarily true. In fact the minimum halo mass that hosts an observed bright galaxy depends on the completeness of the sample. A lower completeness results in a lower minimum halo mass. This does not trivially translate into a higher cross-shot noise, but it would extend the overlap into higher redshifts. For demonstration purposes we assume the sample is complete. In a given redshift bin we perform abundance matching to find $M_{g}^{\rm min}$, using the Sheth-Tormen halo mass function \cite{Sheth:1999mn} and the number density for the BGS survey,  \eqref{eq:nbgs}. The minimum mass is shown in  Figure \ref{fig:Xnoise_mass_range} (left panel), and it compares well with Figure 3.4 of \cite{Aghamousa:2016zmz}.

In order to estimate the maximum HI halo mass, we need to determine the mass range of halos that contribute to HI IM. To this end, we   assume that only halos with circular velocities between 30 and 200 km/s host HI, where \cite{Santos:2015gra}
\be
v_{\rm circ} =30\sqrt{1+z}\left(\frac M{10^{10}M_\odot}\right)^{1/3} ~~{\rm km/s}.
\ee
The maximum HI halo mass is shown in  Figure \ref{fig:Xnoise_mass_range} (left panel), together with the overlap region. The overlap for BGS is only at low redshift.
We do not find any mass range overlap between HI IM and the H$\alpha$ survey. 

 The right panel of  Figure \ref{fig:Xnoise_mass_range} displays the shot-noise power spectra for BGS, HI IM and their cross noise. Our over-estimate of cross-shot noise is well below the galaxy shot noise and rapidly vanishes. Hence we neglect this term in our forecasts. We include the instrumental noise for HI IM, which is clearly much larger than the HI IM shot noise. The step in the instrumental noise  arises from the fact that the frequency ranges of the SKA1 bands and the MeerKAT bands do not perfectly match, as shown in \cite{Fonseca:2019qek}.

\section{The need to add information in non-overlap volumes}\label{apx:comp_MT_allarea}

In table \ref{tab:overlap_V_only} we reproduce table \ref{tab:marginal_gamma} but for the overlap volume only. In both cases, the overlap area is the same, $10^4\deg^2$. The overlap redshift range is $0.1\leq z \leq 0.58$ for the low-$z$ combination and $0.9 \leq z \leq 1.8$ for the high-$z$ combination. At low redshifts, one could in principle use only the overlap area and still obtain a good constraint. At higher redshifts, this is not the case, as most of the information to constrain $\gamma$ comes from the large non-overlap volume.

\begin{table}[!h]
\caption{\label{tab:marginal_gamma_MT} As in Table \ref{tab:marginal_gamma}, but considering only the overlap volumes of  low- and high-$z$ combinations.}  
\centering
\begin{tabular}{llcc}
\\ \hline\hline
  &Survey  &   {\centering $\sigma_{\ln \gamma}\;(\%)$}\\
\hline
Low  redshift &BGS &  5.5 \\
&SKA1 IM2 & 3.8\\
&Combined & 1.9  \\
\hline
High redshift &H$\alpha$ survey  &  4.6 \\ 
&SKA1 IM1 &  12.9\\
&Combined & 4.2   \\
\hline\hline
\label{tab:overlap_V_only}
\end{tabular}
\end{table}

\newpage
\bibliographystyle{JHEP}

\providecommand{\href}[2]{#2}\begingroup\raggedright\endgroup

\end{document}